# Instantaneous effects of photons on electrons in semiconductors

NuoFu Chen[1*], Jiaran Xu[1], Quanli Tao[1], Yiming Bai[1], Jikun Chen[2*]

[1]School of Renewable Energy, North China Electric Power University, Beijing 102206, China.

[2] School of materials, University of Science & Technology Beijing, Beijing 100083, China

*Correspondence and requests for materials should be addressed to N. F. Chen (nfchen@ncepu.edu.cn), and J.K. Chen (jikunchen@ustb.edu.cn)

**Abstract**

It was explored that how photons instantaneously affect electrons in semiconductors comprehensively using quantum mechanics, conventional mechanics, and the theory of relativity. We found that electrons in silicon excited by photons with an energy of 1.12eV, may jump up from top valance band to the bottom of conduction band with an initial speed of $2.543 \times 10^3$ m/s, in $4.977 \times 10^{-17}$ seconds; and affected by photons with energy higher than 4.6eV, the silicon atoms who lose electrons may be catapulted out of crystal in $2.224 \times 10^{-15}$ seconds. These results reasonably explain why laser ablation, laser cutting, and photon-involved rapid thermal annealing are viable processes.

**Keywords:** photon, electron, atom, instantaneous, momentum conservation

The photoelectric effect established by Einstein is well known, which explains how electrons on lower energy levels can jump up to higher levels by absorbing photons, or jump from higher levels to lower levels and release photons[1-3]. However up to this point, how photons act instantaneously on electrons and atoms has not been studied in detail. As a matter of fact, many processes have close correlation with photons, such as laser ablation (LA) and photon-involved rapid thermal annealing (RTA) [4-14]. It is worth the effort to pay more attention to instantaneous effects of photons on electrons and atoms.

In general, it is considered that conventional mechanics is appropriate for microsystem, or life-related environment; quantum mechanics is valid for microsystem; and the theory of relativity is appropriate for huge system, or interstellar space. While, we would like to explore the instantaneous effects of photons of electrons in semiconductors comprehensively using quantum mechanics, conventional mechanics, and the theory of relativity.

Since the introduction of the photon hypothesis by Planck in 1900 [15], the relation between energy E and momentum p of a photon propagating in vacuum has been known to be E = h$\nu$ = cp, where h=6.625×10$^{-34}$ Js is Planck constant, $\nu$ is the frequency, c is the speed of light in vacuum, and p is the momentum of the photon.

The momentum of a photon can be deduced from the fact of light pressure [16,17], i.e.

$$p = h\nu / c. \qquad (1)$$

According to the theory of relativity established by Einstein, the relative mass of a

"particle" can be expressed as [18,19]

$$m = \frac{m_0}{\sqrt{1-\frac{v^2}{c^2}}} \quad (2)$$

where m, $m_0$, and v are the relative mass, rest mass, and speed of the "particle"; c is the speed of light.

On the other hand, it is generally considered that the speed of a photon is the same as the speed of light. There are contradictions and conflicts in this general consideration. If the speed of a photon, v, were the same as the speed of light, c, then the relative mass of the photon would be infinity, according to Eq.2, unless the rest mass of the photon were zero. In this case, the photon's relative mass would be zero too, which would result in the zero momentum of the photon. Obviously, the zero momenta of photons are contradicted by the fact of light pressure, which has been proven by experiments [20-24]. Hereby, there is reasonable to hypothesize that the speed of a photon may not be exactly as the same as the speed of light.

Suppose that the speed of light, c=3.00000000×10$^8$m/s, in Einstein's theory of relativity be unreachable, and the measured speed of a photon be $c_p$=2.99792458×10$^8$m/s in vacuum [25], which is slightly slower than that of light. It is well known that the speed of light slows down in semiconductors, i.e. v=$c_p$/n ( n is refractive index of the semiconductor). Let us take silicon as an example, n=3.87. For a photon with energy of h$v$ =1.12eV, which is the same as the band gap of silicon, Eg. Since h$v$ = m$C_p^2$, so the relative mass of the photon in vacuum is

$$m_{pv} = \frac{hv}{C_p^2} = 1.996 \times 10^{-36}(kg)$$

and the relative mass of the photon in silicon is

$$m_{ps} = \frac{h\nu}{v^2} = \frac{h\nu}{(\frac{C_p}{n})^2} = n^2 m_{pv} = 2.990 \times 10^{-35} (kg)$$

The momentum of a photon in silicon is

$$m_{ps} v_{ps} = n^2 m_{pv} \frac{C_p}{n} = n m_{pv} C_p$$

These results indicate that the relative mass and momentum of a photon in semiconductors will be $n^2$ and n times of the values in vacuum respectively, which is in accordance with Minkowski's description about momenta of light [14], and has been verified in a series of experiments [20-24].

The electrons on top valance band will jump up to the bottom of the conduction band of silicon stimulated by the photons with energy of h$\nu$ =1.12eV, according to Einstein's theory of photoelectric effect [1,2]. However, it has not been mentioned before how the electrons make that jump. Suppose that all the momentum of a photon is absorbed by an electron on top of the valance band, or the conservation of momenta is also satisfied, and the loss in energy during the change is negligible. We may envision that the electron, as shown in Fig.1(a), takes three steps to jump, (i) shifting towards the nucleus with initial speed $v_e$ and bearing the repelling from the nucleus until stable, as shown in Fig.1(b), (ii) accelerated shifting opposite the nucleus until its speed reaches the maximum, -$v_e$, as shown in Fig.1(c), and (iii) decelerated shifting continuously until stable and forming a metastable electron-hole pair or exiton, as shown in Fig.1(d).

In the first step, the electron on top of valance band in silicon will move towards its nucleus with initial speed of $v_0=v_e$, stimulated by the photon. According to the conservation of momentum,

$$m_{ps}v_{ps} = m_e v_e \qquad (3)$$

we obtain the initial speed of electron

$$v_e = \frac{m_{ps}v_{ps}}{m_e} = \frac{2.990 \times 10^{-35} \times 2.998 \times 10^8/3.87}{9.108 \times 10^{-31}} = 2.543 \times 10^3 (ms^{-1})$$

where $m_e=9.108\times10^{-34}$ kg is the rest mass of electrons. The momentum of the photon in silicon is adopted in Eq.3 because most photons with energy of 1.12eV are absorbed inside of the silicon. The electron will be repelled by the atomic nucleus during its movement, until stable. According to Coulomb's law, the repelling force between the electron and nucleus is,

$$F_1 = k\frac{q_1 q_2}{\Delta r_1^2} \approx k\frac{q^2}{\Delta r_1^2} = m_e a_1 \qquad (4)$$

where $k=8.987\times10^9$ Nm²C⁻² is Coulomb's constant, $q_1$ and $q_2$ are the electric charges of the atomic nucleus and the electron; Newton's second law is applied at the end of Eq.4, and $a_1$ is the accelerated speed of the electron. For silicon, suppose that 14 electric charges of the silicon nucleus act with 14 electrons, or each electric charge of the nucleus acts with one electron, so $q_1=q_2=q=1.602\times10^{-19}$ C.

On the other hand, the nucleus applies work, $W_1$, on the electron, and makes the electron's kinetic energy change into potential energy, equaling to the energy of the photon, i.e.

$$W_1 = rF_1 = k\frac{q^2}{\Delta r_1} = h\nu \qquad (5)$$

If the right part of Eq.4 is divided by the square of the right part of Eq.5, Eq.6 can be obtained,

$$a_1 = \frac{(hv)^2}{km_e q^2} = \frac{(hv/q)^2}{km_e} = 1.532 \times 10^{20} \, (ms^{-2}) \qquad (6)$$

The duration, $t_1$, can be calculated from the equation of motion,

$$v_t = v_0 - a_1 t_1 \qquad (7)$$

In this step, $v_0 = v_e$, $v_t = 0$, so

$$t_1 = \frac{v_0}{a_1} = \frac{kq^2 m_e v_e}{(hv)^2} = \frac{km_e v_e}{(hv/q)^2} = 1.659 \times 10^{-17} (s)$$

It can be demonstrated that the durations in the second and third steps are the same as the duration in the first step. Therefore, the whole time for an electron jumping from the top valence band to the bottom of the conduction band is $t = 3t_1 = 4.977 \times 10^{-17}$s. The displacement of the electron jumping from the top valance band to the bottom conduction band is

$$\Delta r = v_e t_1 + \frac{a_1 t_1^2}{2} = 8.435 \times 10^{-14} (m)$$

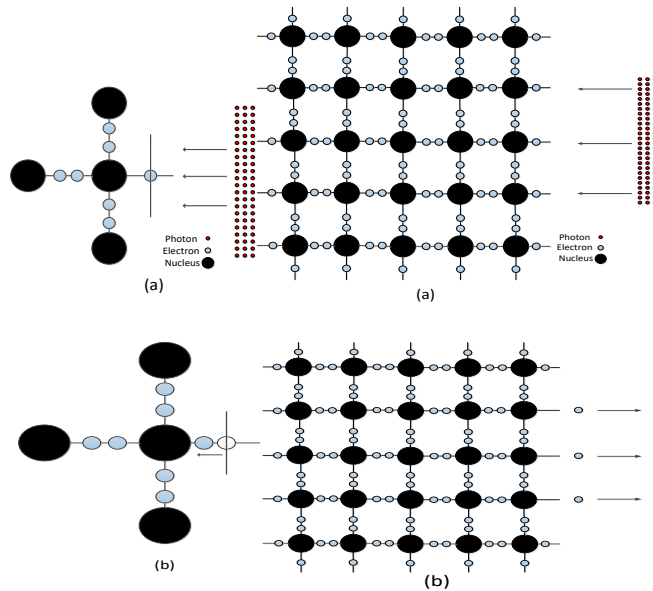

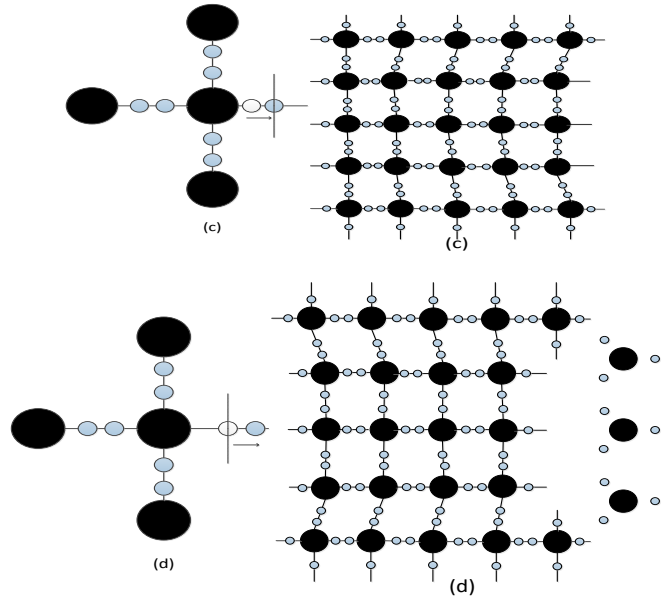

| Figure 1: Sketch maps of (a) the action of photons on valance electrons in silicon, (b) the electron shifted towards its nucleus, (c) the electron shifted opposite its nucleus, and (d) the electron shifted to the bottom of conduction band. | Figure 2: Sketch maps of (a) the action of photons on electrons in surface of silicon, (b) the electrons were ionized, (c) the nuclei moved towards the inner atoms, and (d) the atoms were catapulted out of silicon. |
|---|---|

Figure 2 is helpful to see clearly how the mechanism of laser ablating or cutting works. As shown in Fig.2(a), if the energies of photons are higher than the work function of the semiconductor, the valance electrons of the atoms would be ionized, i.e. the electrons would escape from the semiconductor as shown in Fig.2(b). The work function of silicon is w=4.6eV. With the same analyses as above, we may obtain the duration in the first step of the ionization of the electrons

$$t_{e1} = \frac{v_0}{a} = \frac{km_v c_p}{(hv/q)^2} = 2.542 \times 10^{-19} (s)$$

The momentum of photons in vacuum is adopted here because that most electrons are ionized by the photons in surface of the silicon. The whole time for ionizing the valance electrons is $t_e=3t_{e1}=7.625\times10^{-19}$s.

Once the atoms lose some electrons on one side they will endure counterforce from the escaped electrons, and be attracted by the electrons on the opposite side due to the difference of electric charges. The displacement of the atom would oppress the other atoms near to them as shown in Fig.2(c). The neighbor atoms would take resistant action instead, and catapult the atoms who lose electrons out of the semiconductors as shown in Fig.2(d). Similar to the analyses of the ionization of electrons, the process of de-bonding of silicon atoms takes four steps, (i) accelerated shifting towards the opposite direction of the escaped valance electrons with initial speed $v_a$ until its speed reaching the maximum, $v_{am}$, (ii) decelerated shifting continuously and bearing the repelling from the electrons and neighbor atoms until stable, (iii) accelerated shifting oppositely until its speed reaches the maximum, $-v_{am}$, and (iv) decelerated shifting continuously until stable, and suddenly de-bonding.

The mass of a silicon atom is

$$m_a = \frac{A}{N} = \frac{28.09}{6.025 \times 10^{23}} = 4.662 \times 10^{-23}(g) = 4.662 \times 10^{-26}(kg)$$

where A is the gram atomic weight of silicon, and N is Avogadro constant.

The nucleus will undergo a counter-acting force from the escape of the valance electron, the speed of the nucleus will be

$$v_a = \frac{m_e v_e}{m_a} = \frac{m_{pv} C_p}{m_a} = 1.284 \times 10^{-2}(ms^{-1})$$

On the other hand, the nucleus is attracted by the electrons on the opposite side by force $F_1$, and shifts a displacement of $\Delta r_1$.

$$F_1 = k\frac{q^2}{\Delta r_1^2} = m_a a_1 \tag{8}$$

The electrons apply work, $W_1$, on the nucleus, and make the nucleus move faster,

$$W_1 = \Delta r_1 F_1 = k\frac{q^2}{\Delta r_1} = h\nu \tag{9}$$

Then,

$$a_1 = \frac{(h\nu)^2}{km_a q^2} = \frac{(h\nu/q)^2}{km_a} = \frac{(4.6)^2}{8.987 \times 10^9 \times 4.662 \times 10^{-26}} = 5.05 \times 10^{16}(ms^{-2})$$

If we estimate the duration, $t_{a1}$, with Heisenberg's uncertainty principle,

$$t_{a1} \approx \frac{h}{4\pi h\nu} = \frac{6.625 \times 10^{-34}}{4\pi \times 4.6 \times 1.602 \times 10^{-19}} = 7.154 \times 10^{-17}(s)$$

The speed of the atom, $v_t$, reaches the maximum, $v_{am}$, at $t=t_{a1}$,

$$v_{am} = v_a + a_1 t_{a1} = 3.625(ms^{-1})$$

The displacement, $\Delta r_1$, is

$$\Delta r_1 = v_a t_{a1} + a_1 t_{a1}^2/2 = 9.288 \times 10^{-19}(m)$$

It is reasonable to suppose $t_{a2}=t_{a3}=t_{a4}$, in steps 2-4. The electrons and neighbor atoms would take resistant action with force $F_i$, apply work $W_i$ (i=2-4) to the nucleus, and catapult it out of silicon.

$$F_i = k\frac{Bq^2}{\Delta r_i^2} = m_a a_i \tag{10}$$

$$W_i = k\frac{Bq^2}{\Delta r_i} \tag{11}$$

where B is a coefficient corresponding to the bonding force among the atoms.

Similar as above, we can obtain

$$t_{ai} = \frac{v_{am}}{a_i} = \frac{Bv_{am}}{a_1} = 7.178 \times 10^{-17} B(s)$$

Therefore, the whole duration for an atom in silicon de-bonding is $t= t_{a1}+3t_{a4}+3t_{e1}=7.154\times10^{-17}$s $+2.153\times10^{-16}$Bs $+7.626\times10^{-19}$s.

Suppose B is 10, which means the bonding between atoms is ten times stronger than the bonding between a nucleus and its valance electrons. In this case, the duration for an atom in silicon de-bonding is t=2.224×10⁻¹⁵s.

The threshold value of laser power needed for PLD can be estimated from the density of valance electrons in silicon. The density of atoms in a unit cell of silicon on the projection of plane (001) is $8/a^2$ (a=5.431×10⁻¹⁰ m is the lattice parameter of silicon) [14], which indicate that there are 8 valance electrons in each square of $a^2$. Suppose each photon can act on one valance electron for the ideal situation, the least energy intensity of the photons is

$$I = dh\nu = \frac{8}{(5.431 \times 10^{-10})^2} \times 4.6 \times 1.602 \times 10^{-19} = 19.987 (Jm^{-2})$$

where d is the density of photons, and h$\nu$ =4.6eV is the energy of the photons. The threshold value of a typical UV ns-pulse laser power needed for PLD is

$$P = \frac{I}{\Delta t} = \frac{19.987}{10^{-9}} = 1.999 \times 10^{10} (Wm^{-2}) = 1.999 MW/cm^2$$

In practice, the threshold power density might be of the order of 10–500 MW/cm² for ablation using UV excimer laser pulses of 10 ns duration [8], which is tens to hundreds of the ideal value. The difference between practical and ideal threshold power densities could be comprehended, because no one could recognize how tiny size a photon would be, and one hundredth of the photons may act on an electron.

In conclusion, the affecting process of photons on electrons and furthermore on atoms in semiconductors depends on both the energy and the energy intensity of the photons. The photons whose energies are higher than the band gap of semiconductors can stimulate the valance electrons jumping to conduction bands. Only those photons whose intensity were several ten times higher than the density of the valance electrons and their energies were higher than the work function of the semiconductor could ablate the atoms on the surface of the semiconductor, and make them into plasma. The photons, whose energies are much higher than the band gap and lower than the work function of a semiconductor, can change the positions of atoms in the semiconductor which is responsible for the photon involved rapid thermal annealing.

**Acknowledgements**

Supported by Beijing Natural Science Foundation (2151004), and National Natural Science Foundation (61674013, 51602022).